\documentstyle[b98proc,epsfig]{article}

\newcommand{\ba}{\begin{eqnarray}}
\newcommand{\ea}{\end{eqnarray}}
\input{psfig}

\def\Journal#1#2#3#4{{#1} {\bf #2}, #3 (#4)}

\def\PLB{{\em Phys. Lett.} B}

\def\PRL{\em Phys. Rev. Lett.}

\def\PRD{{\em Phys. Rev.} D}
\def\PRC{{\em Phys. Rev.} C}

\def\ZPA{{\em Z. Phys.} A}

\begin{document}

\title{ELECTROMAGNETIC FORM FACTORS OF BARYONS}

\author{R. BIJKER}
\address{Instituto de Ciencias Nucleares,\\ 
Universidad Nacional Aut\'onoma de M\'exico,\\
A.P. 70-543, 04510 M\'exico D.F., M\'exico}
\author{A. LEVIATAN}
\address{Racah Institute of Physics,\\
The Hebrew University, Jerusalem 91904, Israel}

\maketitle

\abstracts{Nucleon elastic and transition form factors are analyzed in 
the framework of a collective model of baryons. Effects of 
meson cloud couplings and relativistic corrections are considered.}

\section{Introduction}

Electromagnetic form factors of the nucleon and its 
excitations (baryon resonances) provide a powerful tool 
to investigate the structure of the nucleon. 
These form factors can be measured in electroproduction 
as a function of the four-momentum squared $q^2=-Q^2$ of the virtual 
photon. 

In this contribution we present a simultaneous study of the elastic 
form factors of the nucleon and the transition form factors,  
for both of which there exists exciting new data \cite{newdata}.  
The analysis is carried out in the framework of a collective model of 
the nucleon \cite{BIL}. We address the effects of relativistic 
corrections and couplings to the meson cloud. 

\section{Collective model of baryons} 

In the recently introduced collective model of baryons \cite{BIL} 
the radial excitations of the nucleon are described as vibrational 
and rotational excitations of an oblate top. 
%The two fundamental vibrations in this 
%model are associated with the N(1440)$P_{11}$ Roper resonance and the 
%N(1710)$P_{11}$ resonances. The negative parity resonances of the 
%second resonance region are interpreted as rotational excitations. 
%Since each vibrational mode has its own characteristic frequency, 
%there is no problem to explain the relative energy of the Roper 
%resonance with respect to the negative parity resonances. 
The electromagnetic form factors are obtained by folding with a 
distribution of the charge and magnetization over the entire volume 
of the baryon. All calculations are carried out in the Breit frame. 

\subsection{Elastic form factors} 

In the collective model, the electric and magnetic form factors 
of the nucleon are expressed in terms of a common intrinsic 
dipole form factor \cite{BIL}. 
The effects of the meson cloud surrounding the nucleon are taken into 
account by coupling the nucleon form factors to the isoscalar vector 
mesons $\omega$ and $\phi$ and the isovector vector meson $\rho$. 
The coupling is carried out at the level of the Sachs form factors. 
The large width of the $\rho$ meson is taken into account according 
to the prescription of \cite{IJL}, and the constraints from 
perturbative QCD are imposed by scaling $Q^2$ with the strong 
coupling constant \cite{GK}. 

\begin{figure}
\vfill 
\begin{minipage}{.5\linewidth}
\centerline{\epsfig{file=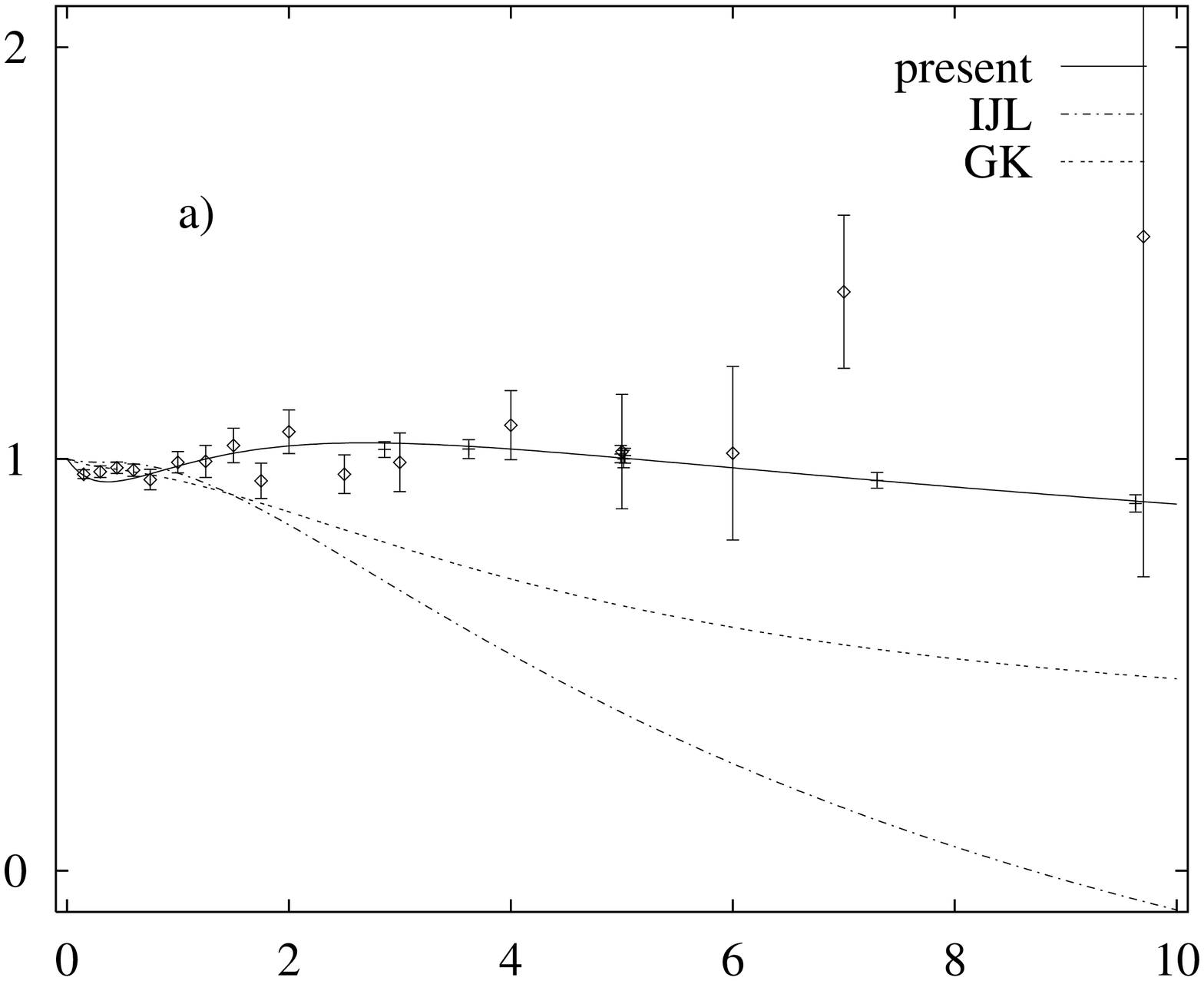,width=\linewidth,angle=-90}}
\end{minipage}\hfill
\begin{minipage}{.5\linewidth}
\centerline{\epsfig{file=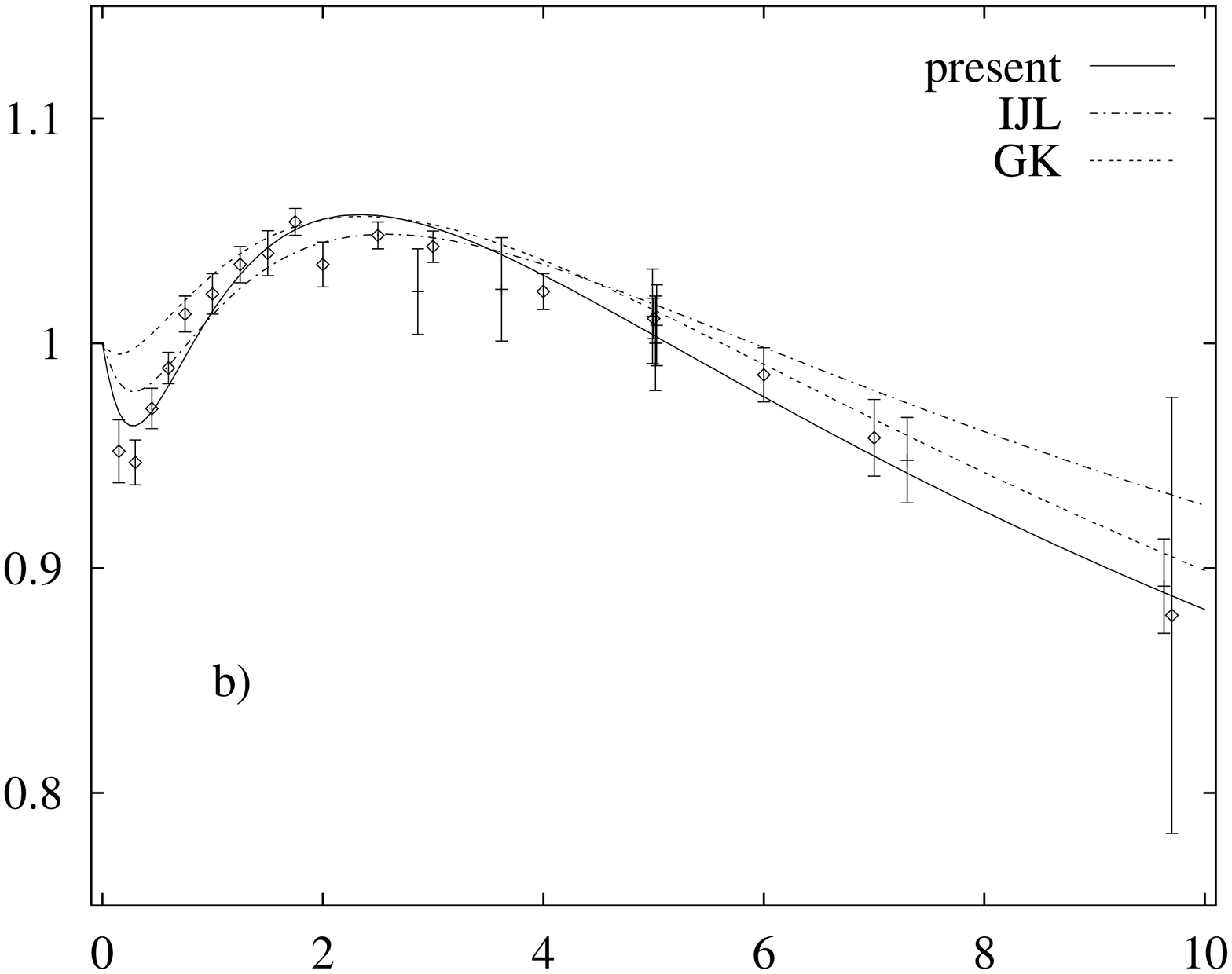,width=\linewidth,angle=-90}}
\end{minipage}
\vspace{1pt}
\begin{minipage}{.5\linewidth}
\centerline{\epsfig{file=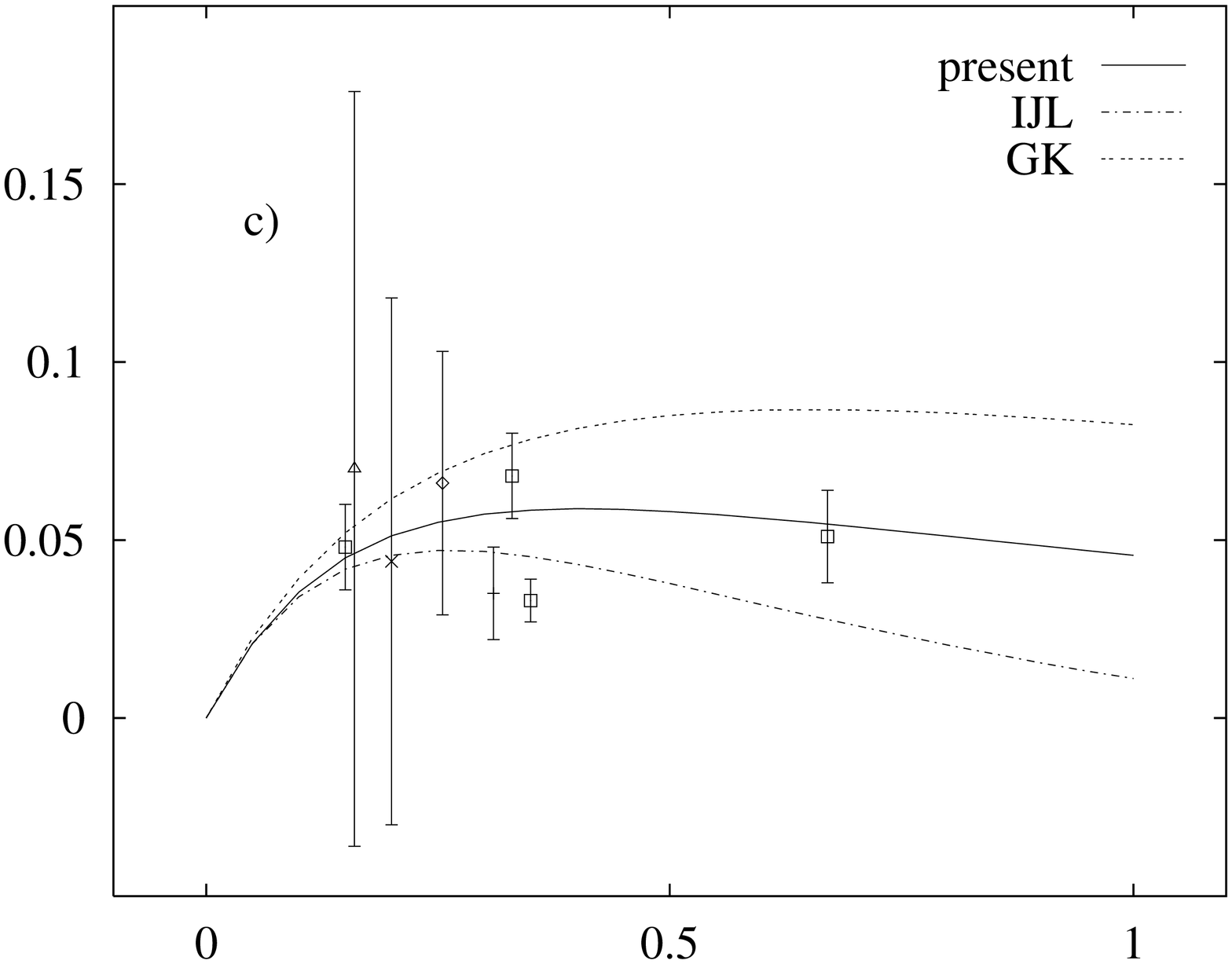,width=\linewidth,angle=-90}}
\end{minipage}\hfill
\begin{minipage}{.5\linewidth}
\centerline{\epsfig{file=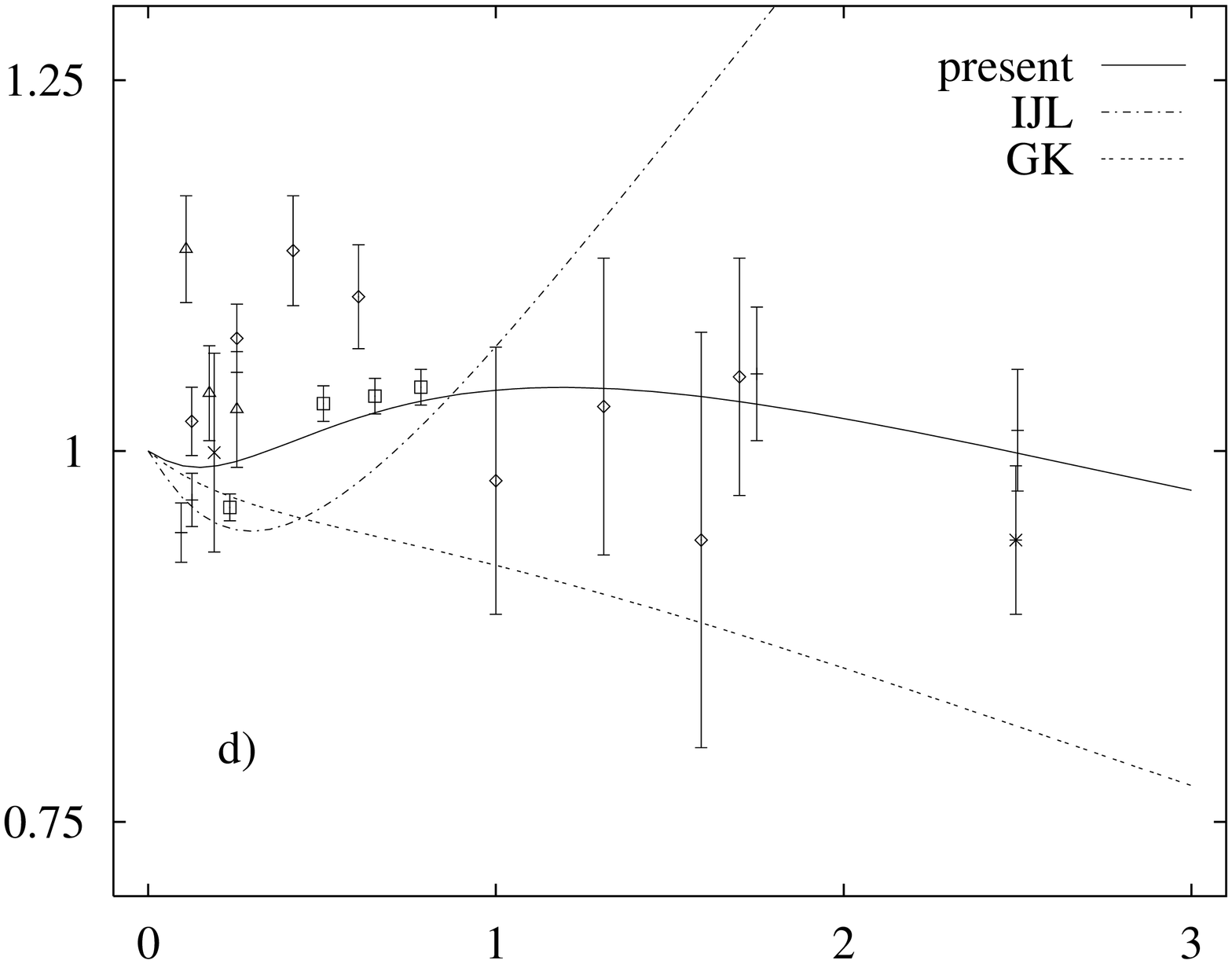,width=\linewidth,angle=-90}}
\end{minipage}
\caption[]{Nucleon form factors as a function of $Q^2$ (GeV/c)$^2$: 
a) $G_E^p/F_D$ from \protect\cite{gemp}, 
b) $G_M^p/\mu_p F_D$ from \protect\cite{gemp}, 
c) $G_E^n$ from \protect\cite{gen} and 
d) $G_M^n/\mu_n F_D$ from \protect\cite{gmn}; 
IJL from \protect\cite{IJL} and GK from \protect\cite{GK};  
$F_D=1/(1+Q^2/0.71)^2$~.}
\label{gempn}
\end{figure}

We carried out a simultaneous fit to all four electromagnetic form 
factors of the nucleon and the nucleon charge radii, and found good 
agreement with the world data, including the new data for the neutron 
form factors (square boxes in Fig.~\ref{gempn}). The oscillations  
around the dipole values are due to the meson cloud couplings. 

\subsection{Transition form factors} 

\begin{figure}[ht]
\centering
\psfig{figure=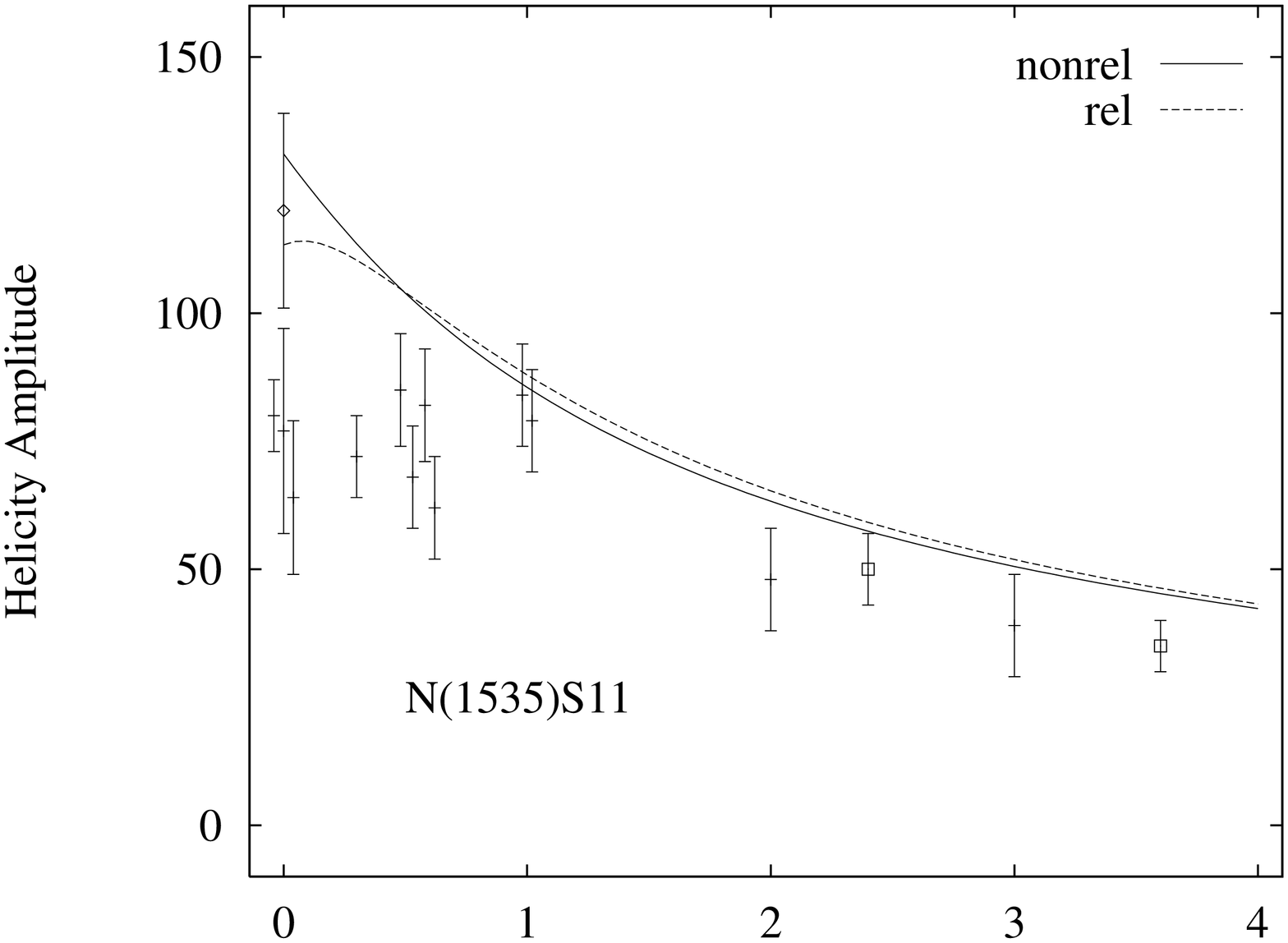,height=7cm,width=10cm,angle=-90}
\caption[]{N(1535)$S_{11}$ proton helicity amplitude in 
$10^{-3}$ GeV$^{-1/2}$ as a function of $Q^2$ (GeV/c)$^2$. 
A factor of $+i$ is suppressed. The data are taken from \cite{s11}.}
\label{n1535}
\end{figure}

Recent experiments on eta-photoproduction have yielded valuable 
new information on the helicity amplitudes of the N(1520)$D_{13}$ 
and N(1535)$S_{11}$ resonances. 
In Table~\ref{ratios} we show the model-independent ratios of 
photocouplings that have been extracted from the new data 
\cite{Nimai1,Nimai2}.
These values are in excellent agreement with those of the collective 
model \cite{BIL}. In the last column we show the effect of relativistic 
corrections to the electromagnetic transition operator. 
In Fig.~\ref{n1535} we show the N(1535)$S_{11}$ proton helicity 
amplitude. The new data are indicated by diamonds and square boxes. 
The effect of relativistic corrections (dashed line) to the 
nonrelativistic results (solid line) in the collective model 
of \cite{BIL} is small, and shows up mainly at small values of $Q^2$. 

\begin{table}
\centering
\caption[]{Ratios of helicity amplitudes.}
\label{helamp} 
\vspace{15pt} 
\begin{tabular}{lllll}
\hline
& & & & \\
N(1535)$S_{11}$ & $A^n_{1/2}/A^p_{1/2}$ 
& --0.84 $\pm$ 0.15 \cite{Nimai1} & --0.81 \cite{BIL} & --0.90 \\
N(1520)$D_{13}$ & $A^p_{3/2}/A^p_{1/2}$ 
& --2.5 $\pm$ 0.2 $\pm$ 0.4 \cite{Nimai2}
& --2.53  \cite{BIL} & --2.66  \\
& & & & \\
\hline
\end{tabular}
\label{ratios}
\end{table}

\section{Summary and conclusions}

We presented a simultaneous analysis of the four elastic 
form factors of the nucleon and the transition form factors 
in a collective model of baryons, and found 
good agreement with the new data presented at this conference. 

The deviations of the nucleon form factors from the dipole 
form were attributed to couplings to the meson cloud 
which were taking into account using vector meson dominance. 
The helicity amplitudes of the N(1520)$D_{13}$ and N(1535)$S_{11}$ 
resonances as well as their model-independent ratios are reproduced 
well. The effect of relativistic corrections is small, and shows up 
mainly at small values of $Q^2$. 

In conclusion, the present analysis of electromagnetic 
couplings shows that the collective model of baryons 
provides a good overall description of the available data.  

\section*{Acknowledgements}

This work is supported in part by DGAPA-UNAM under project IN101997 
and by grant No. 94-00059 from the United States-Israel Binational Science
Foundation (BSF).


\section*{References}
\begin{thebibliography}{99}

\bibitem{newdata}
See e.g. D. Drechsel, D. Rebreyend and P. Stoler, 
these proceedings. 

\bibitem{BIL} 
R. Bijker, F. Iachello and A. Leviatan, 
\Journal{{\em Ann. Phys.} (N.Y.)}{236}{69}{1994}; 
\Journal{\PRC}{54}{1935}{1996}.

\bibitem{IJL}
F. Iachello, A.D. Jackson and A. Lande, 
\Journal{\PLB}{43}{191}{1973}.

\bibitem{GK}
M. Gari and W. Kr\"umpelmann, 
\Journal{\ZPA}{322}{689}{1985}.

\bibitem{gemp}
A.F. Sill et al., 
\Journal{\PRD}{48}{29}{1993}; 
R.C. Walker et al., 
\Journal{\PRD}{49}{5671}{1994}.

\bibitem{gen}
C.E. Jones-Woodward et al., \Journal{\PRC}{44}{R571}{1991};
A.K. Thompson et al., \Journal{\PRL}{68}{2901}{1992};
M. Meyerhoff et al., \Journal{\PLB}{327}{201}{1994};
T. Eden et al., \Journal{\PRC}{50}{R1749}{1994};
H. Schmieden and P. Grabmayr, these proceedings.

\bibitem{gmn}
R.G. Arnold et al., \Journal{\PRL}{61}{806}{1988};
S. Rock et al., \Journal{\PRD}{46}{24}{1992};
A. Lung et al., \Journal{\PRL}{70}{718}{1993};
P. Markowitz et al., \Journal{\PRC}{48}{R5}{1993};
H. Gao et al., \Journal{\PRC}{50}{R546}{1994};
H. Anklin et al., \Journal{\PLB}{336}{313}{1994};
E.E.W. Bruins et al., \Journal{\PRL}{75}{21}{1995};
H. Anklin et al., \Journal{\PLB}{428}{248}{1998}.

\bibitem{Nimai1}
N.C. Mukhopadhyay, J.-F. Zhang and M. Benmerrouche, 
\Journal{\PLB}{364}{1}{1995}.

\bibitem{Nimai2} 
N.C. Mukhopadhyay and N. Mathur, 
preprint nucl-th/9807003.

\bibitem{s11}
V.D. Burkert, \Journal{{\em Int. J. Mod. Phys.} E}{1}{421}{1992}; 
B. Krusche et al., \Journal{\PLB}{397}{171}{1997}; 
C.S. Armstrong et al., nucl-ex/9811001. 

\end{thebibliography}
\end{document}